\documentclass[twocolumn,preprintnumbers,a4,showpacs,prb]{revtex4}
\usepackage{graphicx}%
\usepackage{dcolumn}
\usepackage{bm}
\usepackage{longtable}%

\makeatletter
\def\btt#1{\texttt{\@backslashchar#1}}%
\DeclareRobustCommand\bblash{\btt{\@backslashchar}}%
\makeatother

\begin{document}

\preprint{pd.tex}

\title{Phase diagram of the La$_{1-x}$Ca$_{x}$MnO$_{3}$ compound for $0.5\leq x\leq 0.9$.}
\author{M. Pissas and G. Kallias}
\affiliation{Institute of Materials Science, NCSR,  Demokritos, 153 10 Aghia Paraskevi, Athens, Greece}
\date{\today }

\begin{abstract}
We have studied the phase diagram of La$_{1-x}$Ca$_{x}$MnO$_{3}$ for $0.5\leq x\leq 0.9$ using 
neutron powder diffraction and magnetization measurements. At 300 K all samples are paramagnetic
and single phase with crystallographic symmetry $Pnma$. 
As the temperature is reduced a structural transition is observed which is to a charge-ordered 
state only for certain $x$. On further cooling the material passes to an antiferromagnetic ground
state with Neel temperature $T_N$ that depends on $x$. For $0.8\leq x\leq 0.9$ the structural 
transformation occurs at the same temperature as the magnetic transition.
Overall, the neutron diffraction patterns were explained by considering four phase boundaries for 
which La$_{1-x}$Ca$_x$MnO$_3$ forms a distinct phase: the CE phase at $x=0.5-0.55$, the 
charge-ordered phase at $x=2/3$, the monoclinic and C-type magnetic structure at $x=0.80-0.85$ and 
the G-type magnetic structure at $x=1$. Between these phase boundaries the magnetic reflections 
suggest the existence of mixed compounds containing both phases of the adjacent phase boundaries 
in a ratio determined by the lever rule.
\end{abstract}
\maketitle

The prototypical La$_{1-x}$Ca$_{x}$MnO$_{3}$ compound has been first studied experimentally 
with neutron 
diffraction by Wollan and Koehler\cite{wollan55} and theoretically by Goudenough\cite{goudenough55}
in the middle of 1950's. In the last decade perovskite manganites have attracted 
(for some recent reviews see Ref.~\onlinecite{review} and the references therein)  
significant research interest.
Several works have been published\cite{wollan55,schiffer95,ramirez96,okuda00,heffner01,booth98,%
fujishiro01}
on the phase diagram (or its parts) of
La$_{1-x}$Ca$_{x}$MnO$_{3}$ compounds as function of carrier doping with emphasis around 
$x\sim 1/3$, where the maximum colossal magnetoresistance (CMR) effect has been observed 
(mainly due to its technological relevance).
In the $x\geq 0.5$ regime where the charge/orbital ordering phenomenon appears,
less studies have been devoted.\cite{li99,mori98}
The richness of the phase diagram of La$_{1-x}$Ca$_{x}$MnO$_{3}$ is the result 
of interplay between spin, charge and lattice degrees of freedom.
For $0\leq x \leq 0.21$ the system evolves from the antiferromagnetic $(x=0)$ to 
canted magnetic state and near $x\sim0.2$ towards
the insulating ferromagnetic state.
For the range $0.21< x< 0.5$, a ferromagnetic metallic ground state occurs that undergoes a 
coincident metal-to-insulator and FM-to-paramagnetic transformation on the temperature scale
$150-250$ K. 
In the $0.5\leq x\leq 0.9$  region, the system is not well studied except for some
particular $x$.
Up till now two distinct phases were revealed: $x=1/2$ and $x=2/3$. At half-doping ($x=1/2$), the material
undergoes a phase transition from paramagnetic (insulating) to ferromagnetic-metallic (FM) phase 
at $T_{c}=234$ K and upon further cooling passes to an antiferromagnetic-insulating (AFM) phase at 
$T_{N}^{{\rm c}}=163$ K. The antiferromagnetic phase presents charge and spin ordering, the CE-type structure 
\cite{wollan55}, where real space ordering of Mn$^{3+}$ and Mn$^{4+}$ 
takes place.\cite{radaelli95,radaelli97,chen96,mori98a}
The basic 
characteristic of the $x=1/2$ phase, at low temperatures, is the one after the other ordering of the Mn$^{3+}$ 
and Mn$^{4+}$ ions along the $a\approx \sqrt{2}a_{p}$ directio that leads to a superstructure with propagation vector 
${\bf k}=(1/2,0,0)$. The magnetic structure is a consequence of the structural superstructure.
\cite{radaelli97,kallias98}%
\begin{figure}[htbp] \centering%
\includegraphics[angle=0,width=7.5cm]{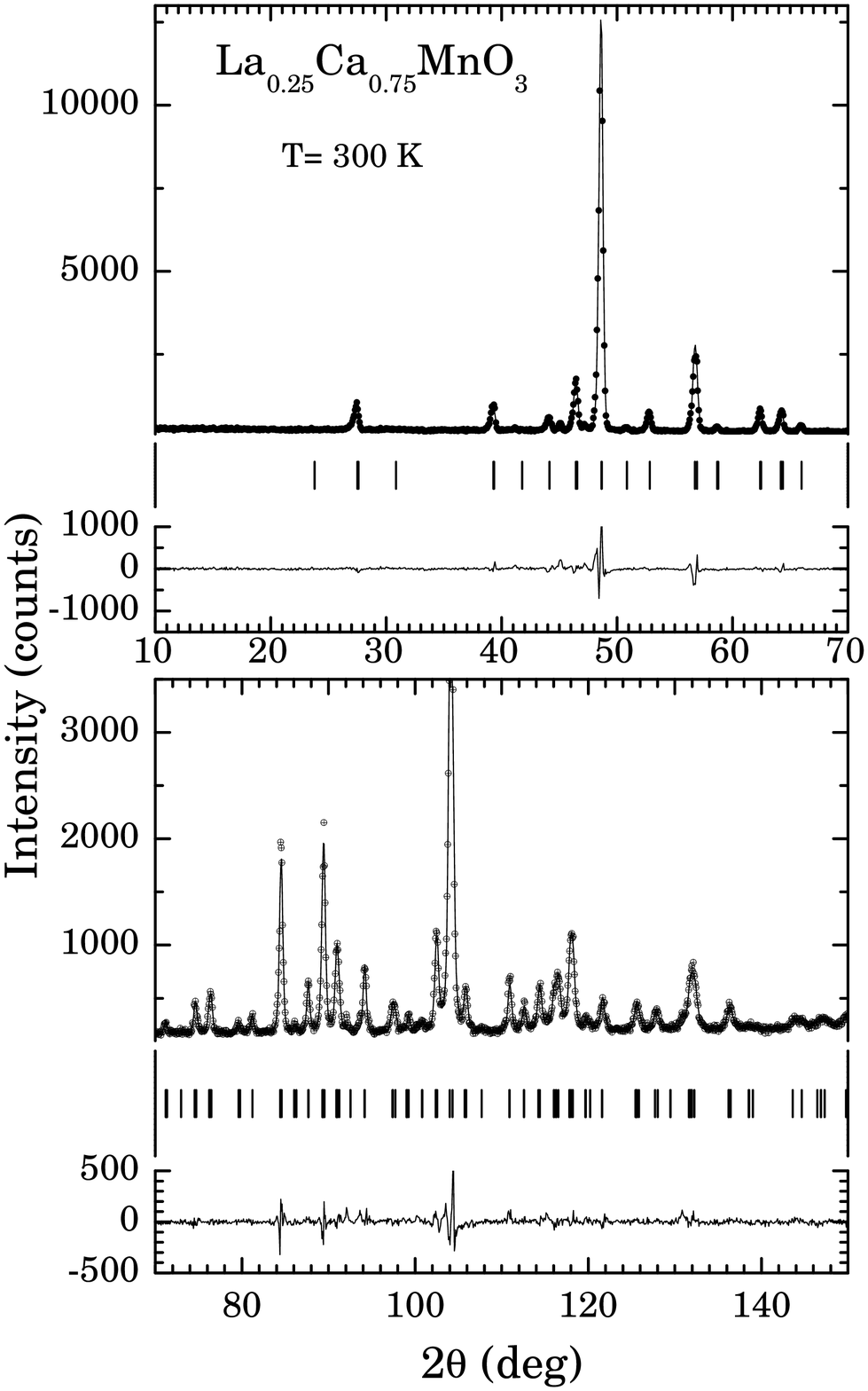}
\caption{
Rietveld refinement pattern for La$_{0.25}$Ca$_{0.75}$MnO$_{3}$ at $T=300$ K ($\lambda=1.798$\AA).
The observed data points are indicated with solid circles, while the calculated pattern is shown as a 
continuous line. The positions of the reflections are indicated with the vertical lines below the pattern.
} 
\label{pattern}%
\end{figure}%
For $x=2/3$, below $T_{{\rm CO}}\approx 260$ K the room temperature phase transforms into a charge-ordered 
low-temperature orthorhombic phase
\cite{ibarra97,fernandez99,radaelli99,wang01,wang00,jo01} 
with space group $Pnma$, but with a tripled unit 
cell ($a_{{\rm CO}}=3a_{{\rm O}},\,b_{{\rm CO}}=b_{{\rm O}}$ and $c_{{\rm CO}}=c_{{\rm O}}$). In addition to 
the charge ordering, the $x=2/3$ compound displays a non-collinear antiferromagnetic structure with the 
$a$ lattice parameter being tripled and the $c$ lattice parameter being doubled with respect to the
average crystallographic unit cell $Pnma$ setting.\cite{ibarra97,fernandez99,radaelli99} 
The crystallographic structure below the charge-ordering temperature is characterized by ordering of the 
$d_{z^{2}}$ orbitals of the Jahn-Teller-distorted Mn$^{3+}$O$_{6}$ octahedra in the 
orthorhombic $ac$ plane and the appearance of superlattice peaks in the x-ray patterns which 
corresponds to the 
tripling of the $a$ axis lattice parameter. The refinement revealed ordering of the Mn$^{3+}$ 
cations in sites as far apart as possible in the $ac$ plane ''Wigner-crystal'' model and 
transverse displacements of the Mn$^{4+}$O$_{6}$ octahedra in the $c$-direction.
Finally, in the $0.9<x<1$ regime both ferromagnetic and antiferromagnetic
interactions\cite{martin99} are present, whereas for $x=1$ CaMnO$_3$ display the G-type  
antiferromagnetic structure.\cite{wollan55,poeppelmeier82}

The charge and orbital ordering can be explained if one considers a delicate balance among two or more 
competing interactions such as Hund's rule coupling, Jahn-Teller distortion and Coulomb interaction.
\cite{hotta01,yunoki00} The CE-state stabilization is induced by the kinetic energy of the $e_{g}$ electron, 
whose motion is restricted by the $t_{2g}$-spin alignment through the double-exchange mechanism.
\begin{table*}[tbp]\centering%
\caption{ Refined unit cell parameters (\AA), fractional atomic positions, anisotropic (for oxygen atoms) 
$b_{ij}\times 10^4$ (\AA$^2$) and isotropic thermal Debye-Waller factors, and reliability
factors derived from neutron diffraction data of La$_{1-x}$Ca$_x$MnO$_3$ samples at $T=300$ K
($x=0.5, 0.55, 0.6, 0.63, 0.72, 0.75$ and $0.9$). 
The space group $Pnma$ (No62) was used for all samples.
La, Ca and O(1) occupy the $4c$\ $(x,1/4,z)$ site, Mn the $4a$\  $(0,0,0)$ site and O(2) the general $8d$ site. 
The numbers in parentheses are estimated standard deviations referring to the last significant digit.}
\label{table1} 
\begin{ruledtabular}
\begin{tabular}{lcccccccc}
$x$ & 0.5 & 0.55 & 0.6 & 0.63 & 0.69 & 0.72 & 0.75 & 0.9 \\ 
\tableline
$a$ & 5.4241(3) & 5.4117(2) & 5.4007(3) & 5.3818(2) & 5.3729(3) & 
5.3628(4) & 5.3535(5) & 5.3082(1) \\ 
$b$ & 7.6446(4) & 7.6154(2) & 7.5916(3) & 7.5683(2) & 7.5612(2) & 7.5559(3)
& 7.5479(5) & 7.4984(2) \\ 
$c$ & 5.4352(4) & 5.4195(2) & 5.4073(3) & 5.3883(2) & 5.3756(3) & 5.3634(4)
& 5.3522(5) & 5.3004(1) \\ 
\tableline
{\bf La/Ca} &  &  &  &  &  &  &  &  \\ 
$x$ & 0.0187(7) & 0.0198(5) & 0.0207(5) & 0.0216(4) & 0.0228(5) & 0.0229(6)& 0.0228(6) & 0.0292(8) \\ 
$z$ &-0.0046(8) & -0.0065(7) & -0.0057(8) & -0.0066(6) & -0.0071(8) & -0.005(1) & -0.007(1) & -0.006(1) \\ 
$B$ & 0.82(5) & 0.84(3) & 0.85(3) & 0.96(3) & 0.84(3) & 0.83(4) & 0.849(4) & 0.92(6) \\
\tableline 
{\bf Mn} &  &  &  &  &  &  &  &  \\ 
$B$ & 0.47(1) & 0.52(4) & 0.46(5) & 0.44(3) & 0.46(3) & 0.42(4) & 0.430(4) & 0.61(6) \\
\tableline 
{\bf O(1)} &  &  &  &  &  &  &  &  \\ 
$x$ & 0.0626(8) & 0.492(1) & 0.4913(9) & 0.4918(7) & 0.4922(7) & 0.4915(9) & 0.4907(9) & 0.4916(8) \\ 
$z$ & 0.5626(8) & 0.06128(56) & 0.0606(6) & 0.0607(5) & 0.0620(5) & 0.0613(8)& 0.0608(9) & 0.062(1) \\ 
$B_{11}$ & 91(9) & 78(5) & 81(6) & 91(4) & 102(6) & 109(8) & 144(11) & 68(16)\\ 
$B_{22}$ & 12(6) & 31(4) & 30(4) & 13(3) & 10(3) & 3(3) & 0(5) & 31(9) \\ 
$B_{33}$ & 91(9) & 78(6) & 81(6) & 91(4) & 102(3) & 109(8) & 144(11) & 23(13)\\ 
\tableline
{\bf O(2)} &  &  &  &  &  &  &  &  \\ 
$x$ & 0.2754(8) & 0.2752(5) & 0.2764(6) & 0.2778(4) & 0.2785(5) & 0.2803(6)& 0.2802(6) & 0.2848(7) \\ 
$y$ & 0.0305(3) & 0.0306(2) & 0.0310(2) & 0.0313(1) & 0.0310(2) & 0.0310(3)& 0.0316(3) & 0.0325(6) \\ 
$z$ & 0.7223(8) & 0.7225(5) & 0.7229(6) & 0.7211(5) & 0.7203(5) & 0.7195(7)& 0.7182(6) & 0.7152(7) \\ 
$B_{11}$ & 66(5) & 74(3) & 76(4) & 81(3) & 77(3) & 81(4) & 84(4) & 22(15) \\ 
$B_{22}$ & 52(5) & 59(3) & 56(3) & 46(2) & 42(2) & 35(3) & 20(3) & 62(7) \\ 
$B_{33}$ & 66(5) & 74(3) & 76(4) & 81(3) & 77(3) & 81(4) & 84(4) & 86(17) \\ 
\tableline
$R_{p}$ & 5.19 & 4.84 & 4.64 & 4.81 & 4.91 & 4.92 & 5.47 & 5.77 \\ 
$R_{wp}$ & 6.76 & 6.22 & 6.42 & 6.34 & 6.36 & 6.37 & 6.98 & 8.11 \\ 
$R_{exp}$ & 4.74 & 4.16 & 4.48 & 3.84 & 3.99 & 4.17 & 3.91 & 3.5 \\ 
$\chi ^{2}$ & 2.03 & 2.24 & 2.06 & 2.72 & 2.54 & 2.33 & 3.19 & 5.35 \\ 
$R_{B}$ & 6.44 & 4.55 & 4.62 & 4.87 & 4.82 & 4.98 & 4.69 & 3.85
\end{tabular}
\end{ruledtabular}
\label{table2}%
\end{table*}%

In the intermediate doping range $1/2<x<2/3$ and for $x>2/3$ the situation is not so clear. 
The only available information is from high resolution electron diffraction lattice images 
\cite{mori98,mori98a,li99} 
which show that charge ordering appears as proper mixtures of the two adjacent distinct commensurate configurations 
according to the lever rule.\cite{mori98,mori98a} 
In addition, the minority phase is not phase-separated into a sizeable area but instead, 
appears as incoherent stacking-fault defects in the otherwise perfect majority phase. 
Above $x\approx 0.8$ a distinct monoclinic phase has been observed \cite{pissas02} with two Mn 
sites which display orbital ordering but no charge ordering.
It is therefore of interest to examine and elucidate the intermediate compositions. 

In the present paper using neutron diffraction and magnetization data we investigated the low
temperature phase diagram of the La$_{1-x}$Ca$_{x}$MnO$_{3}$ compound for $0.5\leq x\leq 0.9$.
The main result is that, in this doping range, La$_{1-x}$Ca$_{x}$MnO$_{3}$ is a distinct phase only for 
certain $x$ ($x=0.5-0.55$, $x=2/3$ and $x=0.80-0.85$) and for the intermediate Ca contents is a mixture of 
the two adjacent boundaries.
\begin{figure}[htbp] \centering%
\includegraphics[angle=0,width=7.5cm]{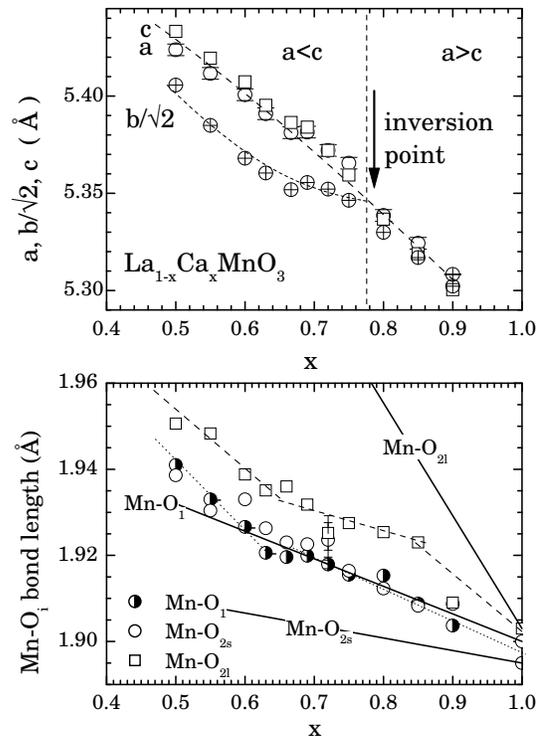}
\caption{
(a) Lattice parameters for the La$_{1-x}$Ca$_x$MnO$_3$ samples at $T=300$ K as obtained from 
Rietveld refinement of neutron powder diffraction data.
(b) Individual Mn-O$_{i}$ ($i=a,p_1,p_2$ denotes bond mainly along $b$-axis, and $a-c$ plane  
respectively) bond lengths at $T=300$ K as a function of $x$. The solid lines 
represent the linearly interpolated bond lengths of the end members. The dash 
and dot lines are guides to the eye. The data for $x=2/3,0.8,0.85$ and 1 are taken
from references \onlinecite{radaelli97,pissas02,poeppelmeier82}.
} 
\label{cell}%
\end{figure}%
La$_{1-x}$Ca$_{x}$MnO$_{3}$ samples with $0.5\leq x\leq 0.9$ were prepared by thoroughly mixing high purity 
stoichiometric amounts of CaCO$_{3}$, La$_{2}$O$_{3}$, and MnO$_{2}$. The mixed powders reacted in air at
temperatures up to 1400$^{\text{o}}$C for several days with intermediate grindings. Finally, the samples were 
slowly cooled to room temperature. Neutron diffraction data were collected on the E6 and E9 diffractometers 
of the research reactor BER II in Berlin. The neutron powder diffraction experiments as a function of 
temperature (in the low angle range) were performed in diffractometer E6 using a wavelength $\lambda =2.44$\AA 
(from the (002) reflection of a pyrolytic graphite monochromator). For crystal structure refinement, data 
were collected on the E9 diffractometer with wavelengths $\lambda =1.798$ and $1.589$\AA (from the (511) and 
(311) reflections of a vertically focusing Ge monochromator), with collimation $\alpha _{1}=10^{/}$
(in pile collimator), 
$\alpha _{2}=20^{/}$(second collimator after monochromator) and $64\times 10^{/}$ collimators in front of 64 
$^{3}$He single detector tubes. The powdered samples were placed in a cylindrical vanadium can ($D=8$ mm) 
which was mounted in an ILL orange cryostat. DC magnetization measurements were performed in a superconducting
quantum interference device (SQUID) magnetometer (Quantum Design).

The structural parameters for all samples at $T=300$ K were refined using the Rietveld method and the 
FULLPROF program. \cite{fullprof} All refinements were carried out in the $Pnma$ space group. In the final 
stage of the refinement procedure the thermal parameters for the oxygen ions were allowed to be anisotropic. 
The final structural parameters for all samples are listed in Table \ref{table1}. As an example, 
figure \ref{pattern} shows the refined pattern for $x=0.75$ at $T=300$ K. In figure \ref{cell}(a)
we plot the variation of lattice parameters as a function of $x$. The cell parameters for the $x=0.5$ 
sample follow the inequality $c>a>b/\sqrt{2}$ which is characteristic of a tolerance factor $t<0.9$. 
As $x$ increases, the lattice parameters $a$ and $c$ decrease almost linearly and with the same slope. 
The $b$ length decreases also but the $b(x)/\sqrt{2}$ curve displays positive curvature and at $x\sim 0.8$ 
merges with the $a(x)$ and $c(x)$ curves. For larger $x$ all lattice parameters continue to decrease with 
nearly equal slopes. 
In addition, even though the difference between $a$ and $c$ is very small, it shows that close to $x=0.8$ a 
change in the inequality relation of the cell parameters occurs and $a>c>b/\sqrt{2}$.

Fig. \ref{cell}(b) depicts the variation of the individual, as well as the 
linearly interpolated Mn-O bond lengths, on $x$. Since all the samples are above the charge ordering
temperature one can expect that the oxygen environment around manganese, in Ca substituted samples, 
acts like a linearly interpolated average of two end members $x=0$ and 1 
(solid lines in Fig. \ref{cell}(b)).
However, our data, clearly show that individual Mn-O bonds do not follow the linearly 
interpolated average.
Furthermore, at certain calcium contain the bond lengths vs $x$ curves change slope abruptly.
These $x$'s are related with $x=2/3$ and 0.85 where phase boundaries occur in the 
La$_{1-x}$Ca$_x$MnO$_3$ phase diagram (vide infra).
It is remarkable that the long bond length of Mn-O$_2$ (which corresponds to Jahn-Teller distortion) 
is far below the expected value for
linearly interpolated of the end members ($d({\rm Mn-O}_2)=2.17(1-x)+1.90x$).
On the other hand the second Mn-O$_{2}$ bond length is larger than the linearly interpolated one. 
Finally the Mn-O$_1$ bond length
follows the interpolated value.
It is logical one to rise the question, what is the explanation for such a behavior?
As has been noticed by Radaelli {\it et al.}\cite{radaelli97b} a small coherent Jahn-Teller
distortion is not incompatible with the large total distortion, provided that the largest part
of it cancels upon application of the space group symmetry operators. The Jahn-Teller distortion
at $T=300 $ K must be truly incoherent, either through static or dynamic disorder.
\begin{figure}[tbp] \centering%
\includegraphics[angle=0,width=7.5cm]{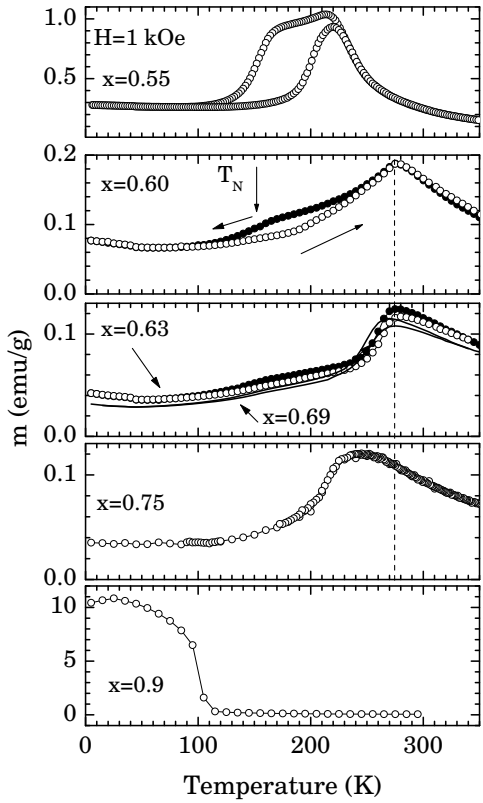}
\caption{
Temperature variation of the magnetic moment per g upon cooling and warming in a 1 kOe magnetic field for
La$_{1-x}$Ca$_x$MnO$_3$ ($x=0.55, 0.6, 0.63, 0.75$ and $0.9$).
} 
\label{squid}%
\end{figure}%

Figure \ref{squid} shows the temperature variation of the magnetic moment under a magnetic field of $1$ kOe 
for samples with $x=0.55$, $0.6$, $0.63$, $0.69$, $0.75$ and $0.9$ as a function of temperature. The 
measurement for $x=0.55$ displays all the characteristics of the $x=0.5$ sample. Namely, the $m(T)$-curve shows 
the well known (for the CE structure) hysteretic behavior at $T_{N}$ which characterizes a first-order 
transition. The measurement for $x=0.6$ reveals remarkable changes with respect to the samples that display 
the CE structure. The peak at high temperatures is related to a structural transition due to charge ordering 
and not to the appearance of ferromagnetic interactions.
At lower temperatures a small anomaly in the $m(T)$ curves ($x=0.6-0.69$) marks the antiferromagnetic 
ordering ($T_N$), in complete agreement with the thermal scans of neutron diffraction data (see below). 
For $0.55<x<0.63$ the heating and cooling $m(T)$-curves display a small hysteresis close to the $T_{N}$. 
Probably this hysteresis is a consequence of the mixed character of the sample (we note that at $T=300$ K 
these samples are single phase material). As the calcium content is further increased the hysteresis and the 
shoulder at $T_{N}$ are further reduced. 
The neutron diffraction data show that for $0.75\leq x\leq 0.85$ the temperature at which additional 
(antiferromagnetic) peaks appear in the neutron diffraction patterns is shifted towards the temperature at 
which the $m(T)$ curves display a peak. 
For $x=0.75$ the structural and the magnetic transition occur with a temperature difference of $20$ K and for 
$0.8\leq x\leq 0.85$ the structural transition (a monoclinic distortion) coincides with the magnetic transition. 
Finally, for the $x=0.9$ sample the $m(T)$-curve shows a variation characteristic of a ferromagnetic phase. 
From this measurement we can not exclude an antiferromagnetic component since its contribution to the bulk 
moment will be smaller by orders of magnitude.
\begin{figure}[tbp] \centering%
\includegraphics[angle=0,width=7.5cm]{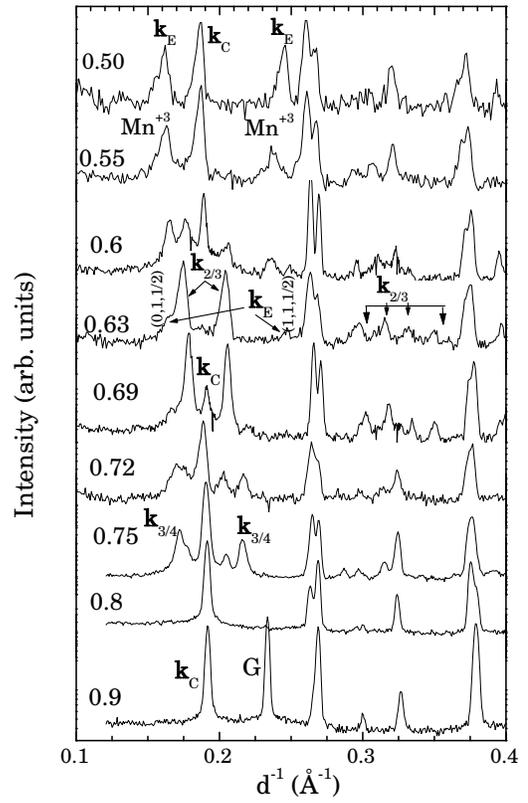}
\caption{
Neutron diffraction patterns for La$_{1-x}$Ca$_x$MnO$_3$ ($x=0.5-0.9$) at $T=2$ K.
} 
\label{fig4}%
\end{figure}%

Figure \ref{fig4} shows the neutron diffraction patterns (NDP) at $T=2$ K for all the samples. 
The pattern for $x=0.5$ displays the classical antiferromagnetic CE structure which can be
described with the propagation vectors ${\bf k}_{{\rm C}}=[1/2,0,1/2]$ and ${\bf k}_{{\rm E}}=[0,0,1/2]$ for 
Mn$^{4+}$ and Mn$^{3+}$ sublattices respectively. The symmetry and the initial orientation of the moments for
the refinement of this structure were taken from the data reported in Ref. \onlinecite{radaelli97}. 
In the CE structure Mn$^{3+}$ and Mn$^{4+}$ are ordered and are located in two crystallographically 
independent sites. The ordered magnetic moments at $T=1.6$ K for the two sublattices were found to be
\ ${\bf m}({\rm Mn}^{3+})=(1.01,0,2.8)$ $\mu _{B}$ and ${\bf m}({\rm Mn}^{4+})=(0,0,2.57)$ $\mu _{B}$, 
respectively. As also noted in Ref. \onlinecite{radaelli97}, the magnetic reflections associated with the 
Mn$^{3+}$ cations are considerably broader than those generated by the Mn$^{4+}$ sublattice. This
behavior has been convincingly explained in this paper by assuming the existence of twinning generated by an 
operation that leaves the Mn$^{4+}$ magnetic sublattice unperturbed while it decreases the size of the 
Mn$^{3+}$ magnetic domains. 

The pattern for the $x=0.55$ sample continues to display the basic characteristics of the CE structure, except 
for the significant broadening of the reflections of the E-type magnetic structure. In addition the $(1,1,1/2)$ 
peak moves to lower angles with respect to the position expected from the crystal structure. These results may 
be explained as due to the smaller size of the Mn$^{3+}$ domains. As the Mn$^{3+}$ domains decrease in size 
with increasing $x$, the relative fraction of surface spins increases and so does the spin disorder in these
domains. Thus, the apparent Mn$^{3+}$ magnetic moment is lowered. Further substitution of La for Ca in the CE 
structure dilutes the Mn$^{+3}$ sublattice with Mn$^{+4}$ ions. This dilution is manifested by broadening and
shifting of the corresponding peaks. Incommensurations, which are doping dependent, may be an 
indication that the antiferromagnetic ground state is in the itinerant limit. By considering AFM in 
the itinerant limit and its competition with CO and ferromagnetism, one can understand the softness of the 
insulating state filling to a magnetic field.\cite{varelogiannis00}
Moreover, the particle-hole asymmetry in the phase diagram of the manganites as well as the CMR phenomenon at 
high temperatures in the hole doping region may also be explained within the same itinerant picture.

\begin{figure}[tbp] \centering%
\includegraphics[angle=0,width=8.5cm]{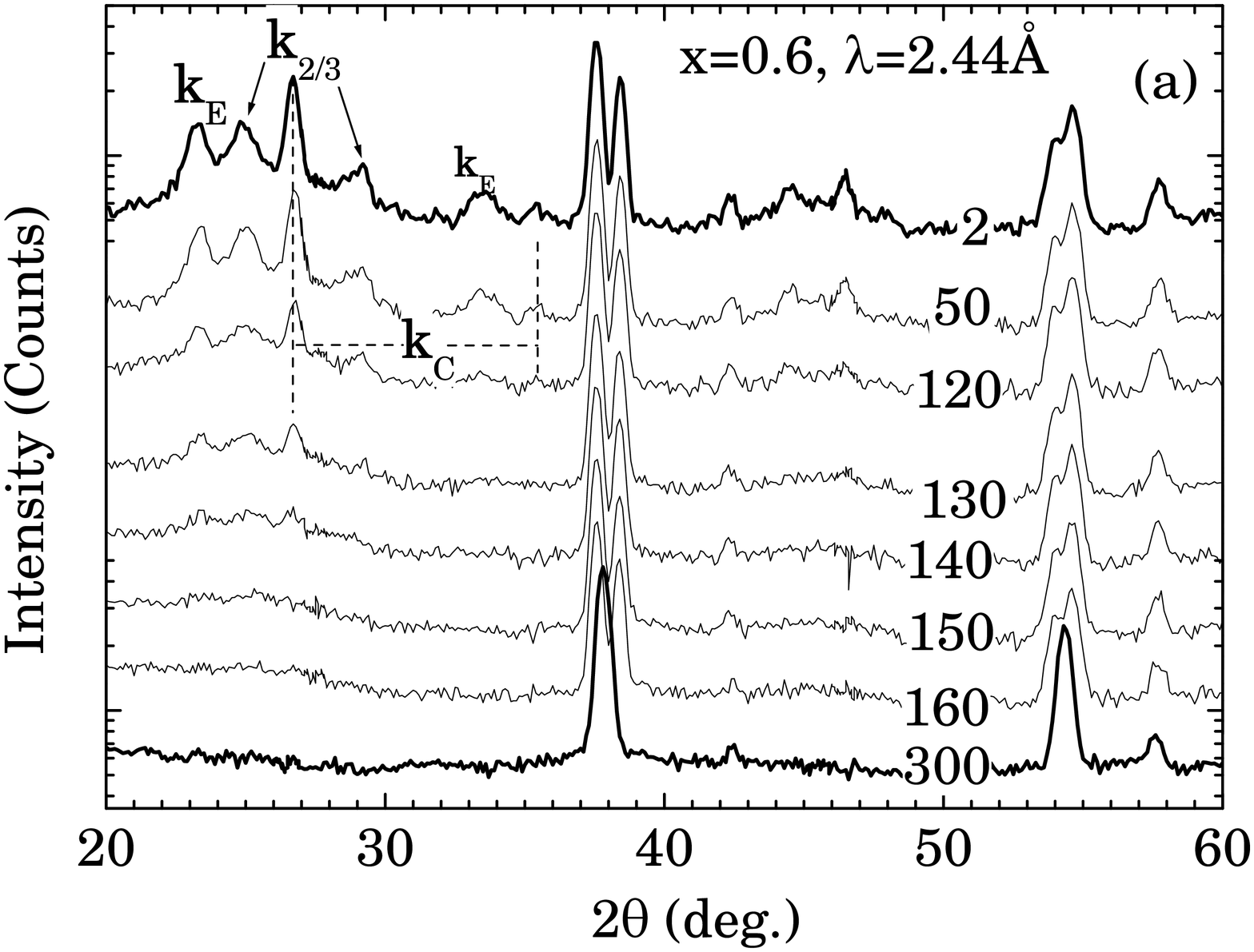}
\includegraphics[angle=0,width=8.5cm]{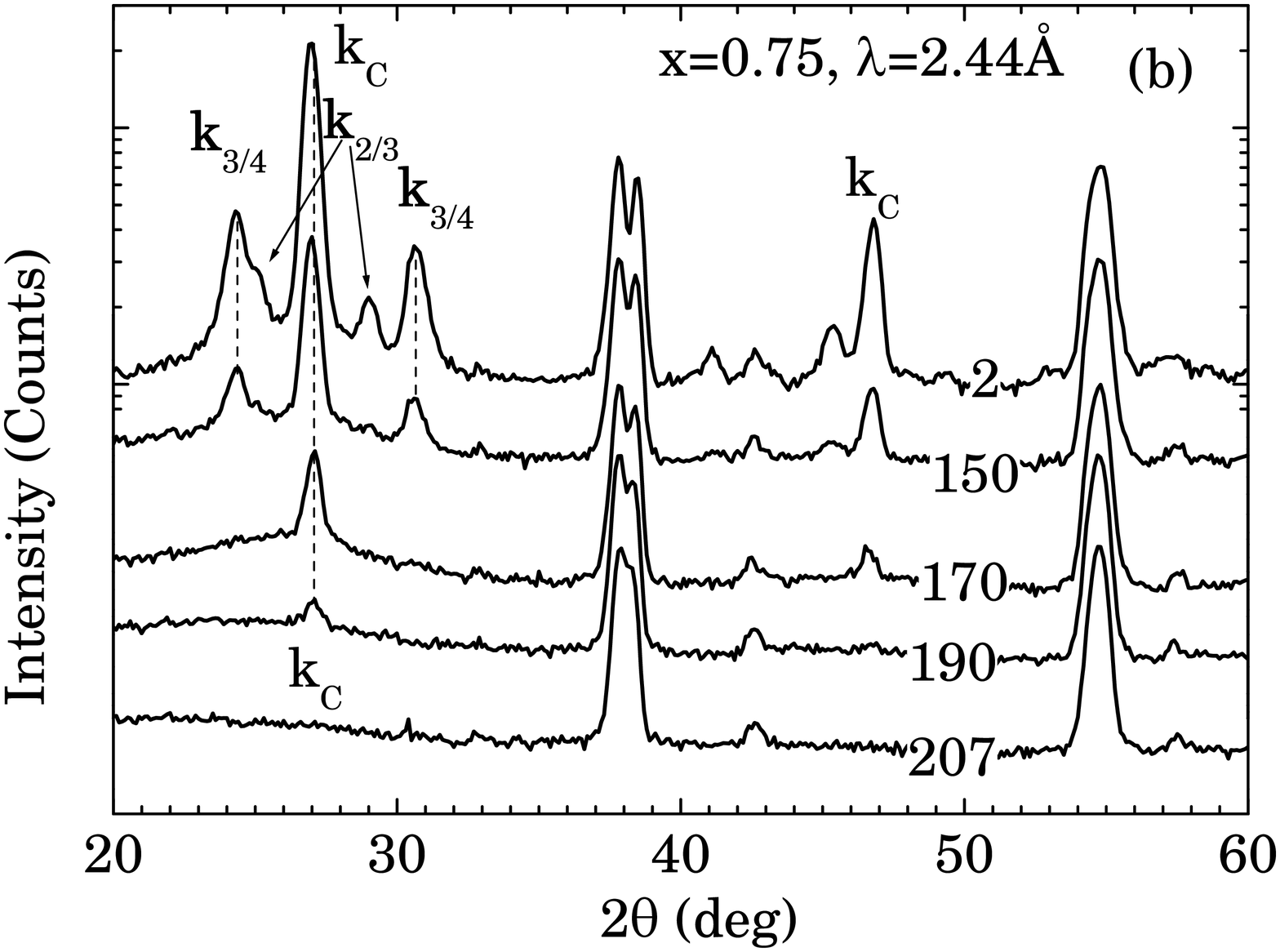}
\caption{
Low angle part of neutron diffraction patterns of La$_{0.4}$Ca$_{0.6}$MnO$_{3}$ (a) and  
La$_{0.15}$Ca$_{0.75}$MnO$_{3}$ (b) for $1.6\le T\le 300$ K.
} 
\label{x06}%
\end{figure}

As $x$ is further increased to $x=0.6$, the NDP show magnetic reflections characteristic of a 
{\em mixed compound}. More specifically, the magnetic peaks of the CE and $x=2/3$ magnetic structures are 
simultaneously present. This finding implies that near this composition, at low temperatures, the system
consists of regions (domains) with CE and $x=2/3$ phases. This may correspond to phase separation, i.e. mixture 
of two phases. We must note that with the term CE we imply a CE phase with a diluted 
Mn$^{+3}$ sublattice. Figure \ref{x06}(a) shows the thermal evolution of NDP's for the $x=0.6$ sample. It is 
interesting to note that the appearance of the magnetic peaks is not characteristic of a second order magnetic 
transition. Rather these broad peaks are reminiscent of short range order. It might be the case that, in this 
regime, the competition between the CE- and $2/3$-structure charge-ordering imposes this complicated behavior.
Close to $x=2/3$ (e.g. $x=0.63$ and $x=0.69$ samples) the NDP are characteristic of 
the antiferromagnetic Wigner crystal charge-ordered structure. 
All the magnetic reflections can be indexed with two propagation vectors: 
${\bf k}_{2/3}=[1/3,0,1/2]$ and 
${\bf k}_{\rm E}=[0,0,1/2]$. Small intensity peaks of the C-type magnetic structure are also present, but these 
peaks should be eliminated exactly at $x=2/3$ or after prolonged annealing that allows the preparation of a
homogeneous sample. This phase is stable in a region around $x=2/3$. 
A Rietveld refinement of the
magnetic and crystal structure at $T=2$ K for $x=0.63$ sample using the 
Wigner crystal charge-ordered
model\cite{radaelli99} gave ordered-magnetic moment for 
two sublattices
${\bf m}({\rm Mn}^{+3})=(1.99,0,1.24)$ $\mu_B$  and
${\bf m}({\rm Mn}^{+4})=(1.87,0,1.58)$ $\mu_B$, respectively.

For $x>2/3$ the Mn$^{+3}$ sublattice is further diluted, thereby leading to the development of the C-type 
magnetic structure. 
Magnetic peaks which correspond to the propagation vector ${\bf k}_{3/4}=[1/4,0,1/2]$ 
appear. At $x=0.75$ one can expect a new charge-ordered phase with $a=4a_{Pnma}$. Our NDP for this calcium 
concentration revealed coexistence of magnetic peaks corresponding to the $x=2/3$ phase as well as to the 
propagation vectors ${\bf k}_{3/4}$ and ${\bf k}_{\rm C}$. In order to check if prolonged annealing does stabilize
a phase with $a=4a_{Pnma}$, two samples were measured. The first was annealed at 1400$^{\rm o}$C for $7$ days and 
the second for $14$ days. Even though the intensity of the ${\bf k}_{3/4}$ peaks increases, the peaks remain 
broad. On the other hand, the ${\bf k}_{2/3} $ peaks reduce in intensity but do not disappear. 
Remarkable is the disappearance of the magnetic peaks which correspond to the
propagation vector ${\bf k}_{2/3}$ around 150 K. The corresponding peaks of ${\bf k}_{3/4}$
are disappeared at higher temperature, whereas those of ${\bf k}_{\rm C}$ in the interval 170-190 K
(see Fig. \ref{x06}(b)).

For $0.8\leq x\leq 0.85$ a monoclinic distortion and antiferromagnetic C-type long range order are observed.
\begin{figure*}[tbp] \centering%
\includegraphics[angle=0,width=14cm]{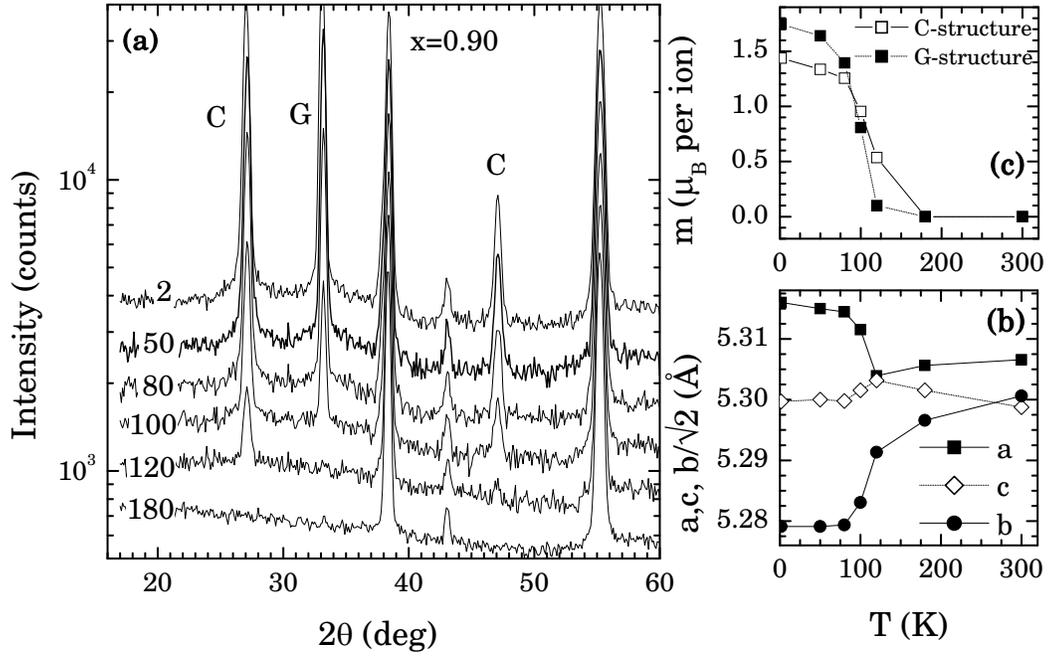}
\caption{(a) Neutron diffraction patterns of La$_{0.1}$Ca$_{0.9}$MnO$_{3}$ for $2\leq T\leq 180$ K. 
(b) Temperature variation of the lattice parameters. 
(c) Temperature variation of the ordered magnetic moment of the C- ang G-type magnetic structures.
} 
\label{x09}%
\end{figure*}%
\begin{figure*}[tbp] \centering%
\includegraphics[angle=0,width=14cm]{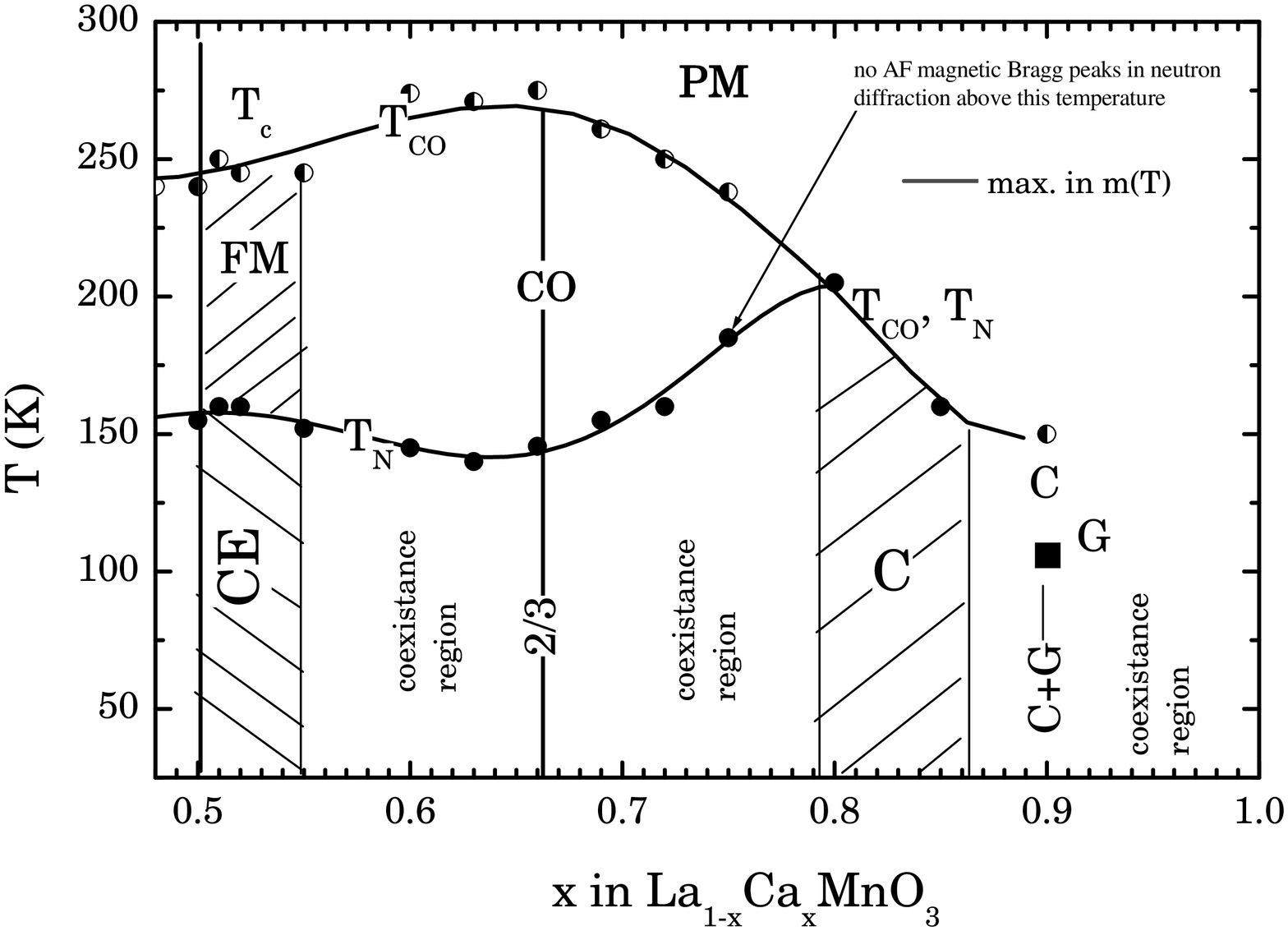}
\caption{
Phase diagram of the La$_{1-x}$Ca$_x$MnO$_3$ compound in the high Ca doping region $x=0.5-0.9$.
} 
\label{pd}%
\end{figure*}%
Figure \ref{x09}(a) depicts the thermal evolution of the NDPs for $x=0.9$. The NDPs show a mixed magnetic 
structure comprising both C- and G-type magnetic peaks. 
The presence of two propagation vectors may imply that the magnetic structure is non-collinear. 
The variation of the magnetization shows ferromagnetic behavior with $T_{c}\sim 110$ K. 
We note that the peaks corresponding to the C and G magnetic structures do not disappear simultaneously. 
The C-type peaks are present at $T=120$ K, whereas the G-type peaks disappear in agreement 
with the magnetization measurement. No anomaly was observed in the magnetization measurements since the 
large ferromagnetic signal should mask the much smaller antiferromagnetic one. The ferromagnetic moment is
favorable for the non-collinear magnetic structure. 
Regarding the crystal structure, the monoclinic distortion observed for the samples $x=0.8$ and $x=0.85$ 
disappears or is below the resolution limit of the instrument. At the temperature where the magnetic peaks 
appear, the $b$-axis decreases and $a$-axis increases, while the $c$-axis remains nearly constant 
(see fig. \ref{x09}(b)). Rietveld refinement at $T=1.6$ K gave ordered magnetic moments 
${\bf m}_{\rm C}=(1.81(5),0,0)$ and ${\bf m}_{\rm G}=(2.34(5),0,0.3(1))$ $\mu _{B}$ for the C- and G-type magnetic 
structures, respectively. Figure \ref{x09}(c) shows the temperature variation of the ${\bf m}_{\rm C}$ and 
${\bf m}_{\rm G}$ ordered moments for $x=0.9$.

In Fig. \ref{pd} we compiled both the magnetization and neutron data to produce a phase diagram for the 
La$_{1-x}$Ca$_{x}$MnO$_{3}$ system in the doping region $0.5\leq x\leq 0.9$. This is the main result of our
work.
On increasing the Ca content above the $x\sim 0.5$ boundary, the system retains the CE-structure even though 
the Mn$^{+3}$ sublattice is diluted with Mn$^{+4}$ ions. In the doping range $0.55<x<2/3$ the CE and 
$2/3$ phases coexist due to the competition between the forces which stabilize these phases. It is only when 
the Mn$^{+3}/$Mn$^{+4}$ ratio is a certain rational number that charge-ordering occurs. Otherwise, the system 
consists of a mixture of the two phases according to the lever rule. Another clear phase boundary occurs 
in the vicinity of $x=0.8$. However, this phase does not clearly display charge-ordering. \cite{pissas02}
In the regime $2/3<x<0.8$ there is coexistence of the two phases with an additional complication due to the 
fluctuations which tend to stabilize $x=3/4$ charge-ordered state. Finally, for $0.85<x<1$ the 
C- and G-type magnetic structures coexist (G is the magnetic structure adopted by the CaMnO$_{3}$). 

The temperature where the structural transition occurs seems to follow the lever rule (fig. \ref{pd}), 
i.e. it increases in the region $0.55<x<2/3$ towards the value $T_{\rm CO}(x=2/3)\simeq 270$ K. The corresponding 
Neel temperature (we must not forget that in this region we have a mixture) is approximately constant or 
decreases slightly. Perhaps this is due to the fact that the $T_N$ temperatures of the CE and $2/3$-structures 
are approximately equal. Similarly, the reduction of the structural transition temperature can be explained 
with the lever rule, 
in agreement with the high resolution electron diffraction data.\cite{mori98} 
At the region where only the C-structure is present the structural and the magnetic 
transition occur concurrently. The temperature where the magnetic peaks appear ($T_N$) increases approaching 
the value of the $x=0.8$ sample. In several previous works $T_{N}$ was shown to decrease monotonically up to 
$x=0.9$, in contrast with our neutron data.

In conclusion, we studied the low temperature phase diagram of the La$_{1-x}$Ca$_{x}$MnO$_{3}$ compound for 
$0.5\leq x\leq 0.9$ using elastic neutron diffraction data and magnetic measurements. The neutron diffraction 
patterns can be consistently explained by considering the existence of four phase boundaries $x=0.5-0.55$, 
$x=2/3$, $x=0.80-0.85$ and $x=1$ at which the material forms a distinct crystallographic and magnetic phase.
At the intermediate $x$ it consists of a mixture of the two adjacent boundaries according to the lever rule.
The variation of the $T_N$ and of the structural transition temperature with the Ca doping content can be 
constructed by applying the lever rule. A final interesting result is that for $x=0.80-0.90$ the magnetic and 
the structural transition occur simultaneously.

\acknowledgments
Partial support for this work was provided by the E.C. through the CHRX-CT93-0116 Human Capital and Mobility 
Program.


\begin{references}
\bibitem{wollan55}  E. O. Wollan and W. C. Koehler, Phys. Rev. {\bf 100}, 545 (1955).

\bibitem{goudenough55}  J. B. Goudenough, Phys. Rev. {\bf 100}, 564 (1955).

\bibitem{review}
{\it Colossal Magnetoresistive Oxides}, Gordon and Breach Science Publishers,Edited by Y. Tokura (2000) ;
{\it Colossal Magnetoresistance, Charge Ordering and Related Properties of Manganese Oxides},World Scientific, Edited by C. N. Rao and B. Raveau (1998);
{\it Physics of Manganites} Kluwer Academic/Plenum Publishers, Edited by T. A. Kaplan and S. D. Mahanti (1999);
N. R. Rao, Anthony Arulraj, A. K. Cheetham, and Bernard Raveau, J. Phys.: Condens. Matter {\bf 12},R83 (2000);
A. P. Ramirez,  J. Phys.: Condens. Matter {\bf 9}, 8171 (1997); 
Y. Tokura and Y. Tomioka, J. Magn. Magn. Mater. {\bf 200}, 1 (1999); 
J. M. D. Coey, M. Viret, and S. von Molnar, Adv. Phys. {\bf 48}, 167 (1999);
M. B. Salamon and M. Jaime, Rev. of Modern Phys., {\bf 73}, 583 (2001);
E. Dagotto, T. Hotta, A. Morea,  Physics Reports {\bf 344}, 1 (2001);
E. L. Nagaev, Physics Reports,{\bf 346}, 387 (2001).

\bibitem{schiffer95}  P. Schiffer, A. P. Ramirez, W. Bao and S. -W. Cheong,
\prl {\bf 75}, 3336 (1995).
\bibitem{ramirez96}  A. P. Ramirez, P. Schiffer, S. -W. Cheong, C. H. Chen, W. Bao, T. T. Palstra, 
P. L. Gammel, D. J. Bishop and B. Zegarski, 
\prl {\bf 76}, 3188 (1996).
\bibitem{okuda00} T. Okuda, Y. Tomioka, A. Asamitsu, and Y. Tokura,
\prb {\bf 61}, 8009 (2000).
\bibitem{heffner01} R. H. Heffner, J. E. Sonier, D. E. MacLaughlin, G. J. Nieuwenhuys,
G. M. Luke, Y. J. Uemura, William Ratcliff II, S-W. Cheong, G. Balakrishnan,
\prb {\bf 63}, 094408 (2001).
\bibitem{booth98} C. H. Booth, F. Bridges, G. H. Kwei, J. M. Lawrence, A. L. Cornelius
 and J. J. Neumeier, \prb {\bf 57}, 10440 (1998). 
\bibitem{fujishiro01} H. Fujishiro, T. Fukase, M. Ikebe,
J. Ohys. Soc. Jpn {\bf 70}, 628 (2001).
\bibitem{li99} J. Q. Li, M. Uehara, C. Tsuruta, Y. Matsui and Z. X. Zhao,
\prl {\bf 82},2386 (1999).
\bibitem{mori98}  S. Mori, C. H. Chen and S.-W.Cheong, Nature {\bf 392}, 473 (1998).




\bibitem{radaelli95}  P. G. Radaelli, D. E. Cox, M. Marezio, S-W. Cheong, 
P. E. Schiffer and A. P. Ramirez, \prl {\bf 75}, 4488 (1995).



\bibitem{chen96}  C. H. Chen and S-W. Cheong, Phys. Rev. Lett. {\bf 76}, 4042 (1996).
\bibitem{mori98a}  S. Mori, C. H. Chen, and S-W. Cheong, Phys. Rev. Lett. {\bf 81}, 3972 (1998).
\bibitem{radaelli97}  P. G. Radaelli, D. E. Cox, M. Marezio and S-W. Cheong, 
\prb {\bf 55}, 3015 (1997). 

\bibitem{kallias98} G. Kallias, M. Pissas, and A. Hoser, Physica B {\bf 276-278}, 778 (2000)



\bibitem{ibarra97} M. R. Ibarra, J. M. De Teressssa, J. Blasco, P. A. Algarabel,
C. Marquina, J. Garcia, J. Stankiewicz and C. Ritter, 
\prb {\bf 56}, 8252 (1997).
\bibitem{fernandez99}  M. T. Fernandez-Diaz, J. L. Martinez, J. M. Alonso, and E. Herrero, 
\prb {\bf 59}, 1277 (1999).
\bibitem{radaelli99}  P. G. Radaelli, D. E. Cox, L. Capogna, S-W. Cheong, and M. Marezio, 
\prb {\bf 59}, 14440, (1999).
\bibitem{wang01}  R. Wang J. Gui, Y. Zhu and A. R. Moodenbaugh, 
\prb {\bf 63}, 144106 (2001).
\bibitem{wang00} R. Wang J. Gui, Y. Zhu and A. R. Moodenbaugh, 
\prb {\bf 61}, 11946 (2000);
\bibitem{jo01} Younghun Jo, J.-G Park, Chang Seop Hong, N. H. Hur and H. C. Ri, 
\prb {\bf 63}, 172413 (2001).



\bibitem{martin99} C. Martin, A. Maignan, M. Hervieu and B. Raveau
Phys. Rev. B {\bf 60}, 12191 (1999).
\bibitem{poeppelmeier82} K. R. Poeppelmeier, M. E. Leonowicz, J. C. Scanlon, J. M. Longo, 
and W. B. Yelon, Journal of Solid State Chemistry, {\bf 45}, 71 (1982).
















\bibitem{hotta01}  T. Hotta, E. Dagotto, H. Koizumi, and Y. Takada, Phys. Rev. Lett. {\bf 86,} 2478 (2001).

\bibitem{yunoki00}  S. Yunoki, T. Hotta, and E. Dagotto, Phys. Rev. Lett. {\bf 84}, 3714 (2000).

\bibitem{pissas02}  M. Pissas, G. Kallias, M. Hofmann and D. M. T\"{o}bbens
\prb (to be published).

\bibitem{fullprof}  J. Rodr\'{\i}guez-Carvajal, Physica B {\bf 192}, 55 (1993).

\bibitem{radaelli97b}  P. G. Radaelli, G. Iannone, M. Marezio, H. Y. Hwang, S-W. Cheong, 
J. D. Jorgensen and D. N. Argyriou,
\prb {\bf 56}, 8265 (1997).
\bibitem{varelogiannis00}  G. Varelogiannis, Phys. Rev. Lett. {\bf 85}, 4172 (2000).

\end{references}
\end{document}